\documentclass[
    ,final            % use final for the camera ready runs
    ,numberedheadings % uncomment this option for numbered sections
  ]
  {aipproc}
\layoutstyle{6x9}
\newcommand{\be}{\begin{eqnarray}}
\newcommand{\ee}{\end{eqnarray}}

\newcommand{\pp}{\pi^+ \pi^-}
\newcommand{\cosine}{$\langle cos\theta \rangle$ \,}

\title{Hard Exclusive Electroproduction of Two Pions off 
       Proton and Deuteron at HERMES}

\author{Pasquale di Nezza$^1$, \underline{Riccardo Fabbri}$^2$\\
        ({\it on behalf of the HERMES Collaboration})}{
address={$^1$INFN, Laboratori Nazionali di Frascati, Via Enrico Fermi 40,
         00044 Frascati, Italy\\
         $^1$E\_mail: Pasquale.DiNezza@lnf.infn.it\\
         $^2$Department of Physics, Ferrara University, 
         Via delle Scienze 12, 44100 Ferrara, Italy\\
         $^2$E\_mail: rfabbri@fe.infn.it}
}

\begin{document}
\begin{abstract} 
Exclusive electroproduction of $\pi^+\pi^-$ 
pairs off hydrogen and deuterium targets 
has been studied with the HERMES experiment. 
The angular distribution of the $\pi^+$ in 
the $\pi^+\pi^-$ rest system has been studied 
in the invariant mass range 
$0.3 < m_{\pi\pi} <1.5$ GeV.
Theoretical models derived in the framework of the 
Generalized Parton Distributions show that 
this angular distribution receives 
only contributions from the 
interference between the isoscalar channel 
$I=0$ and the isovector channel $I=1$.
\end{abstract}

\maketitle

%%%%%%%%%%%%%%%%%%%%%%%%%%%%%%%%%%%%%%%%%%%%
%% INTRODUCTION
%%%%%%%%%%%%%%%%%%%%%%%%%%%%%%%%%%%%%%%%%%%%
\small

\vspace{-1.5cm}
\section{Introduction}

\begin{figure}[b]
  \includegraphics[height=3.5cm,width=3.8cm]{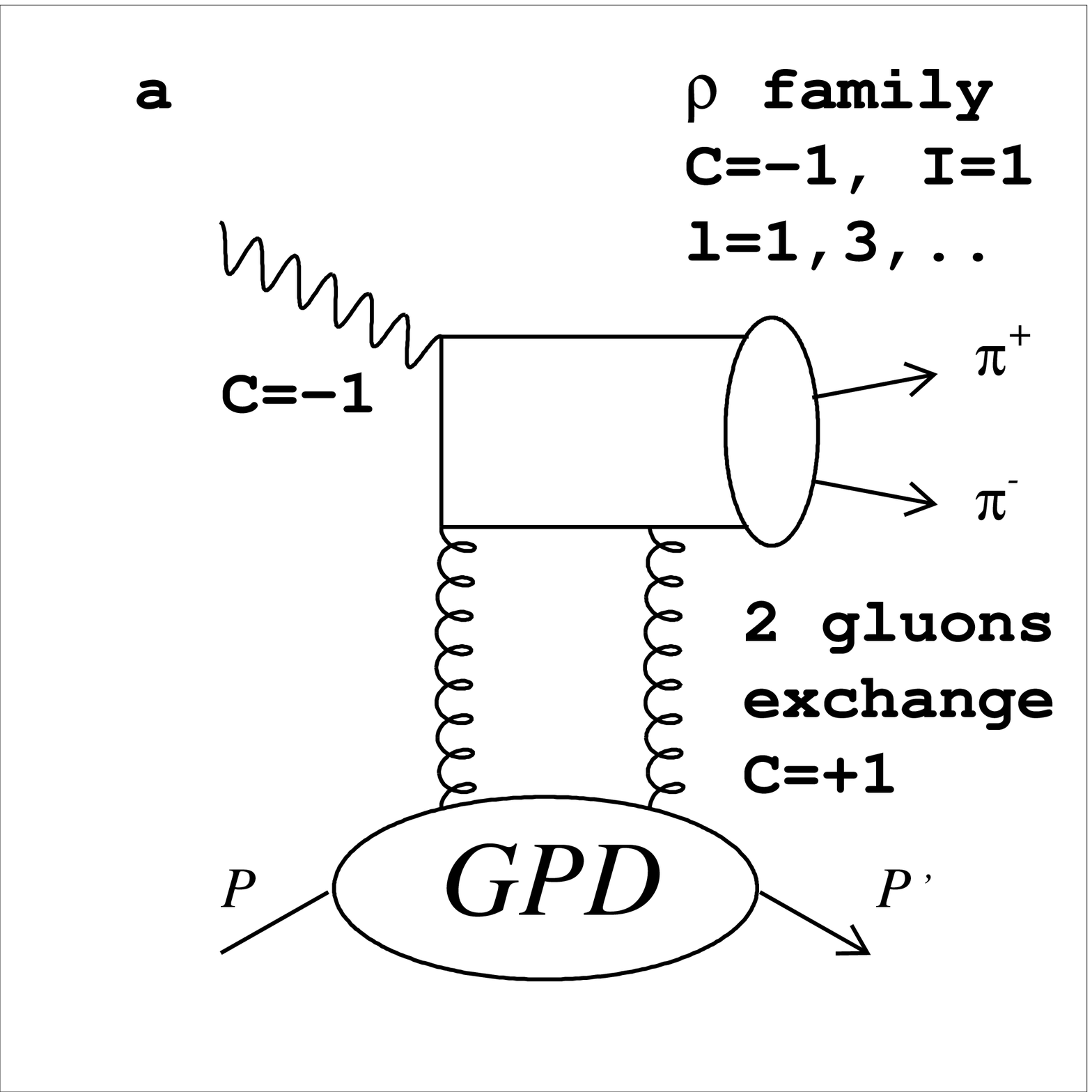} 
  \includegraphics[height=3.5cm,width=3.8cm]{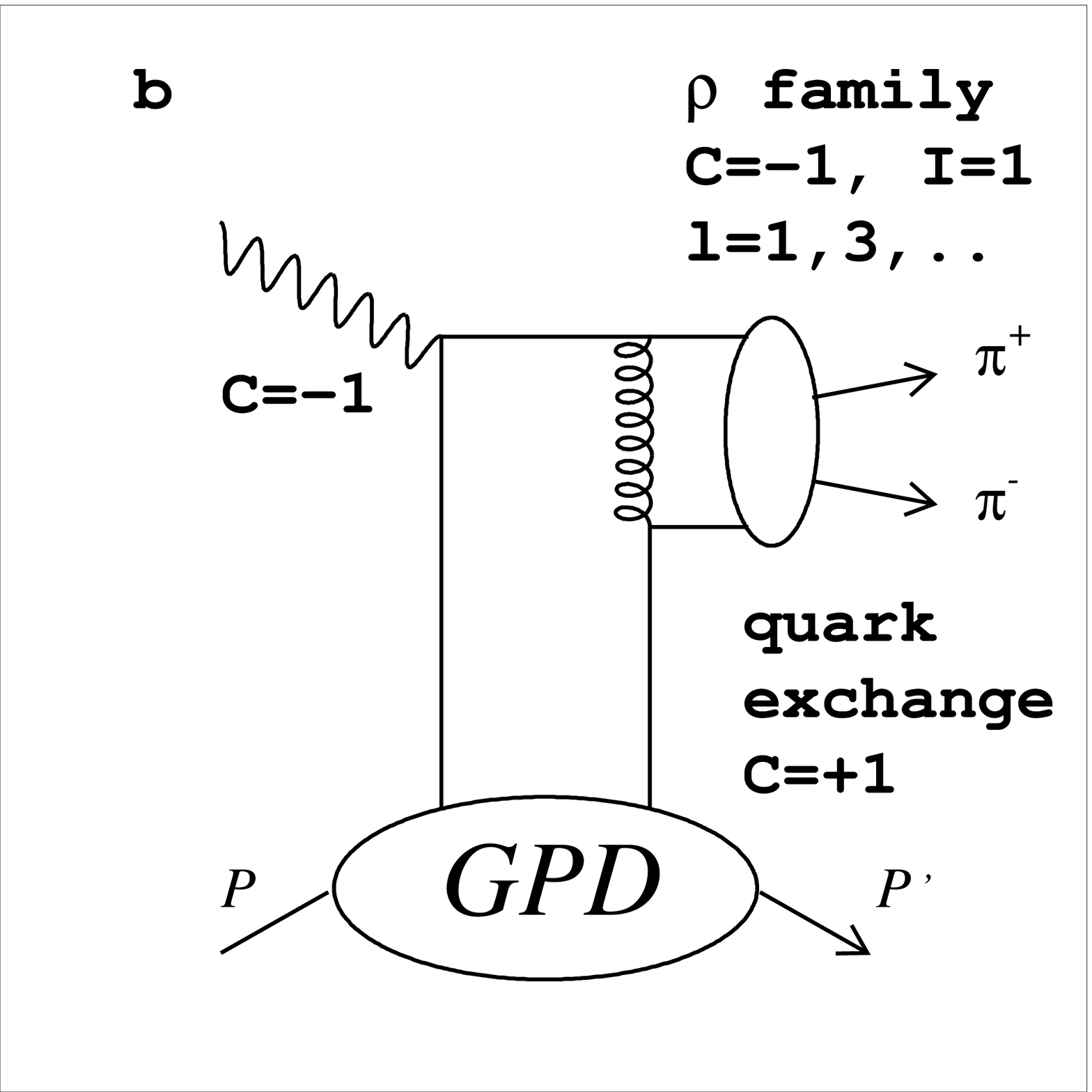} 
  \includegraphics[height=3.5cm,width=3.8cm]{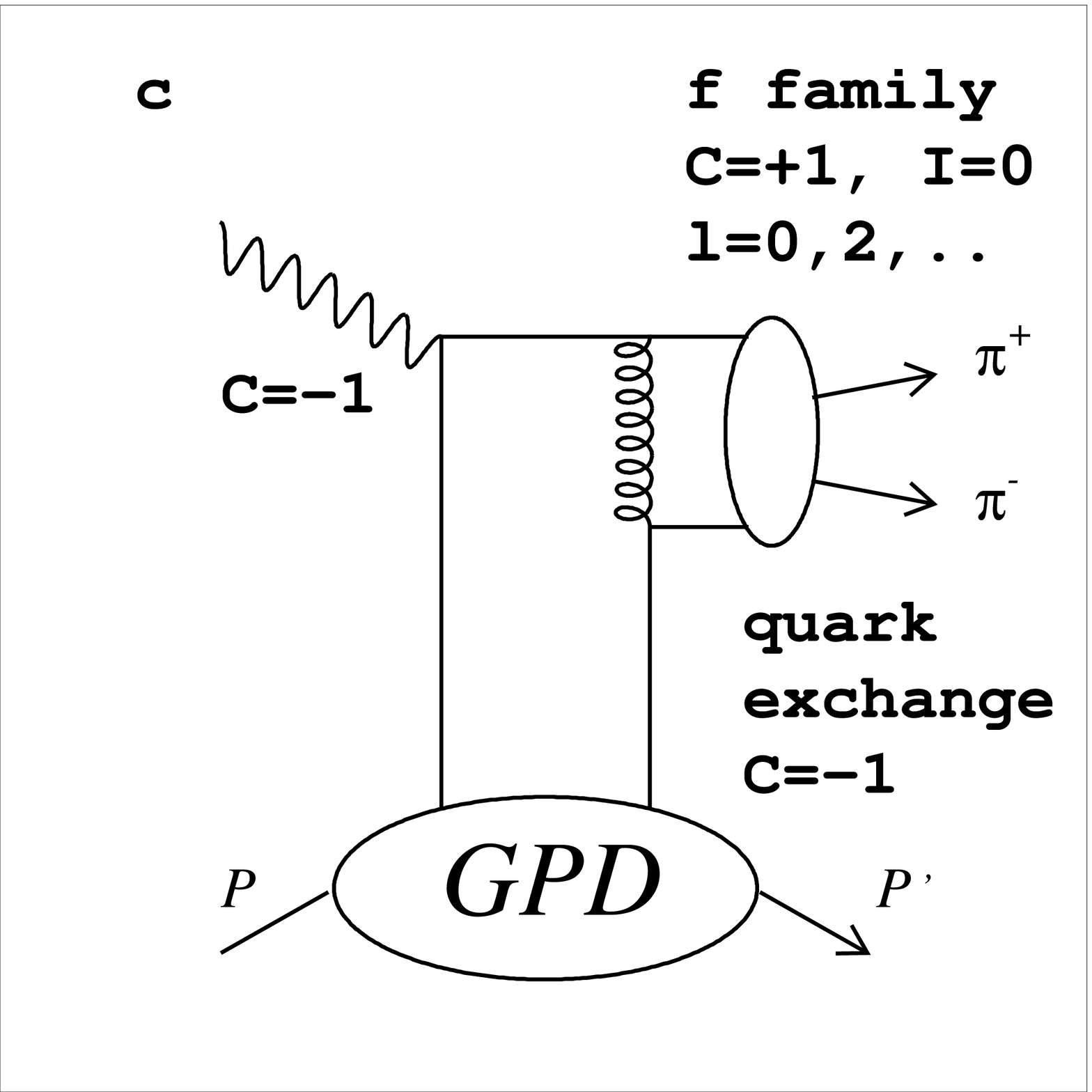}  
  \includegraphics[height=3.5cm,width=3.8cm]{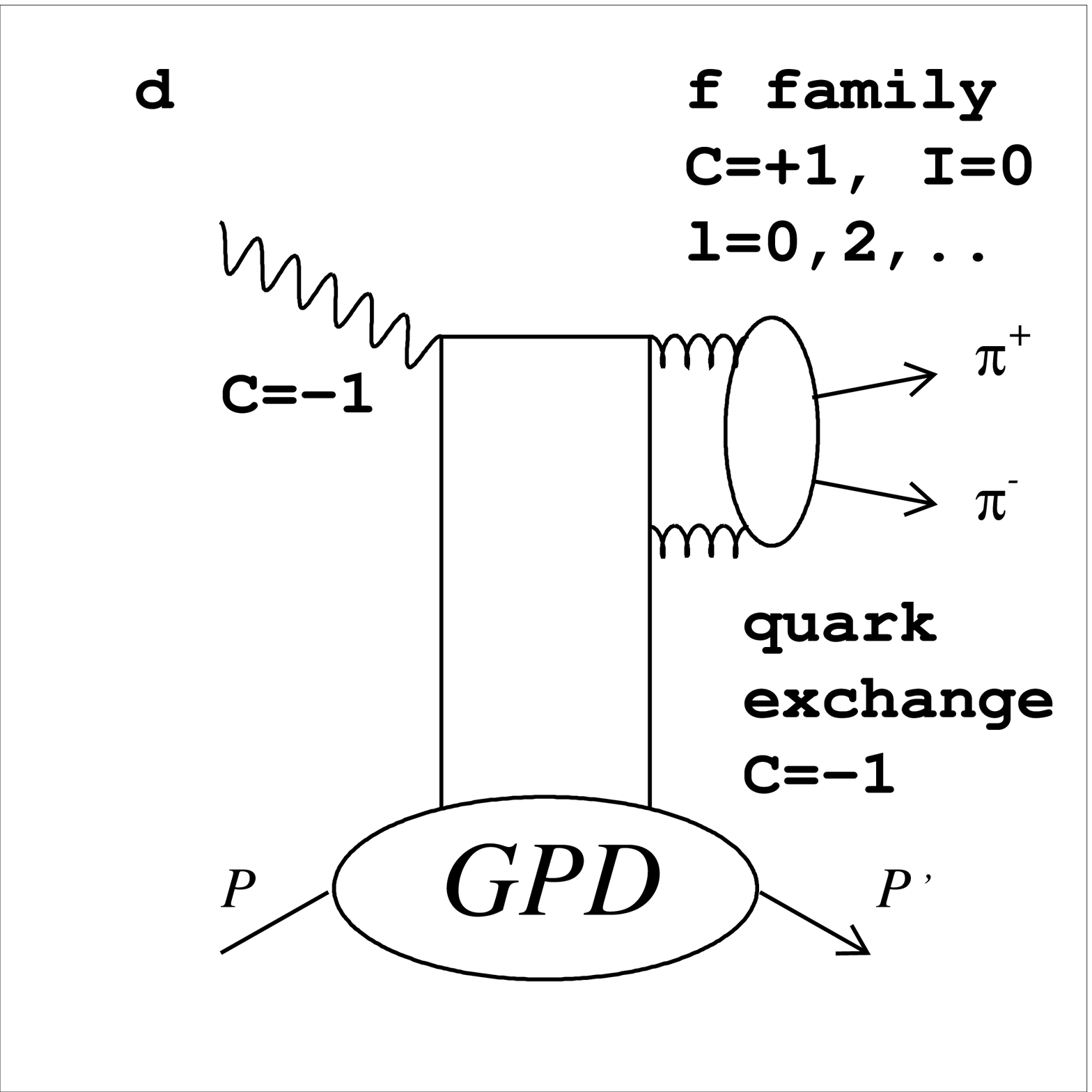} 
  \caption{Leading twist diagrams for the hard exclusive
           reaction $e^+ p \rightarrow e^+ p \; \pi^+ \pi^-$.
           The gluon exchange ({\it a)} gives rise to pions
           in the isovector state only, while 
           the quark exchange ({\it b,c,d}) gives rise to 
           pions both in the isoscalar and in the isovector state.}
  \label{fig:feynman}
\end{figure}

\vspace{-0.3cm}
The analysis of hard exclusive production
of $\pi^+\pi^-$ pairs off unpolarized targets 
of hydrogen and deuterium at HERMES is presented. 
Recent theoretical studies \cite{collins,polyakov00,polyakov01} 
have shown that the exclusive process 
$ e^+ p \longrightarrow e^+ p \; \pi^+\pi^- $
can be described in the framework of 
Generalized Parton Distributions ({\it GPD}s) 
\cite{radyushkin,ji98,ji97}.
The diagrams relevant for this reaction 
at leading twist are shown in Fig. \ref{fig:feynman}.
The pion pairs may be produced through gluon or quark 
exchange with the target, either from quark 
(Fig. \ref{fig:feynman}a,b,c) or from gluon fragmentation
(Fig. \ref{fig:feynman}d).
In the range of the considered $\pi^+\pi^-$ invariant mass 
($m_{\pi\pi}$), 
both resonant and non-resonant contributions are present.
The particular state describing the  
$\pi^+\pi^-$ pair, with the angular momentum quantum number 
{\it l} being odd or even, also defines unequivocally 
the quantum numbers C-Parity ({\it C}) and Strong Isospin 
({\it I}) of this state.
At small values of the Bjorken variable $x_{Bj}$, 
pions are produced mostly 
in the isovector state, because the dominant 
mechanism is two-gluon exchange with 
positive C-parity (Fig. \ref{fig:feynman}a) \cite{polyakov99}.
At large $x_{Bj}$ the production of pion pairs 
is dominated by $q\bar{q}$ exchange 
(Fig. \ref{fig:feynman}b,c,d) \cite{diehl}, which leads 
to a sizable admixture of pion pairs with isospin 
zero.

In the HERMES kinematics the $\pi^+\pi^-$ 
cross section is dominated by the isovector 
channel, i.e. $\rho^0$ production. 
The less copious isoscalar $\pi^+\pi^-$ 
production can be investigated by studying the 
interference between the odd and even {\it l} 
wave production. 
An observable suitable to probe this 
interference appears to be the {\it intensity density} 
\cite{polyakov00,polyakov01},  
defined as the $l^{th}$ Legendre Polynomial $P_l(cos\theta)$ 
moment 
\be
  \langle P_l(cos\theta)\rangle^{\pp} =
  \frac{\int_{-1}^1 d cos\theta \, P_l(cos\theta)\,
  \frac{d\sigma^{\pp}}{d cos\theta}}
  {\int_{-1}^1 d cos\theta\,\frac{d \sigma^{\pp}}{d cos\theta}},
\ee
where $\theta$ is the scattering angle 
of $\pi^+$ meson in the $\pi^+\pi^-$ rest frame 
with respect to the direction of the recoiled 
system \cite{diehl}.
Some of the above defined intensities 
are non-zero only if the interference 
between the isoscalar and the isovector 
channels is present.
Experimentally they are obtained as average 
values of $P_l(cos\theta)$. 
In this paper we describe the measurement of 
the intensity density for $l=1$ which is  
calculated by the average $\langle cos\theta \rangle$. 
%

%======================
%==   DATA ANALYSIS  ==
%======================
%
%-------------------Data-analysis---------------%
%
\vspace{-0.5cm}
\section{Data analysis}

\vspace{-0.3cm}
The data have been accumulated with the 
HERMES forward spectrometer \cite{hermes} during the 
running period 1996-2000 of HERA. 
The 27.6 GeV positron beam was scattered off  
hydrogen and deuterium targets, respectively. 
Events have been selected with one 
positron track and  two oppositely charged hadron  
tracks ($E_h > 1.0$ GeV) without additional neutral 
clusters in the calorimeter. 
The exclusivity of these events was ensured by 
imposing a further cut on the inelasticity 
$\Delta E = \frac{M^2_X - M^2_P}{2 M_P}$, 
where $M_X$ is the invariant mass of the undetected
system and $M_p$ is the proton mass.   
As explained above, to enhance the isoscalar production 
and consequently the interference between the isovector 
and isoscalar channel, it was also required 
that $x_{Bj} > 0.1$. 
Moreover, in order to enter the {\it hard} regime of the 
process, constraints on $Q^2$ ($Q^2>1$ GeV$^2$) and  
$W$ ($W>2$ GeV) were imposed, where $W$ is the invariant mass 
of the virtual-photon nucleon  system.

\begin{figure}[t]
  \includegraphics[height=.3\textheight]{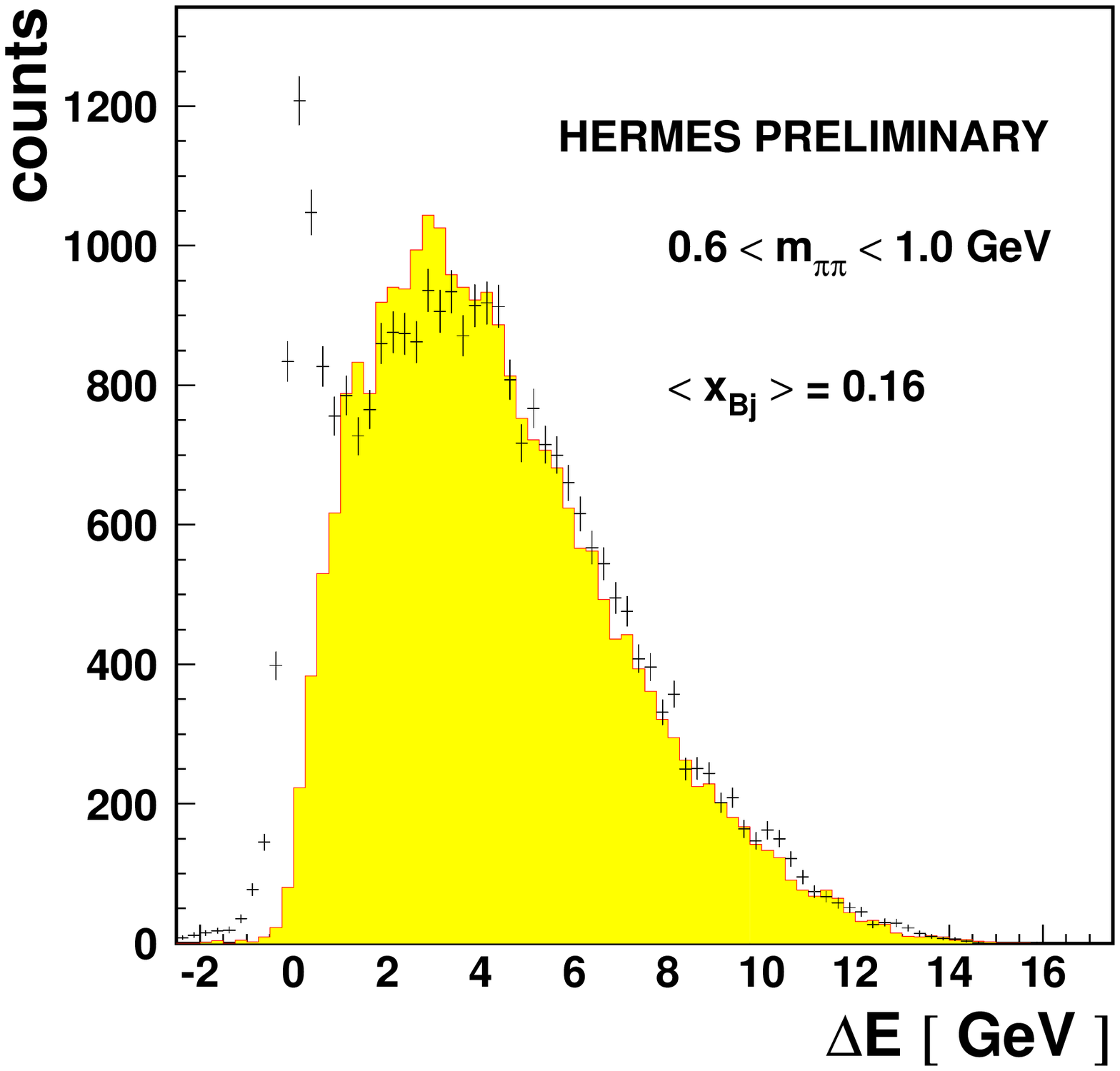}
  \includegraphics[height=.3\textheight]{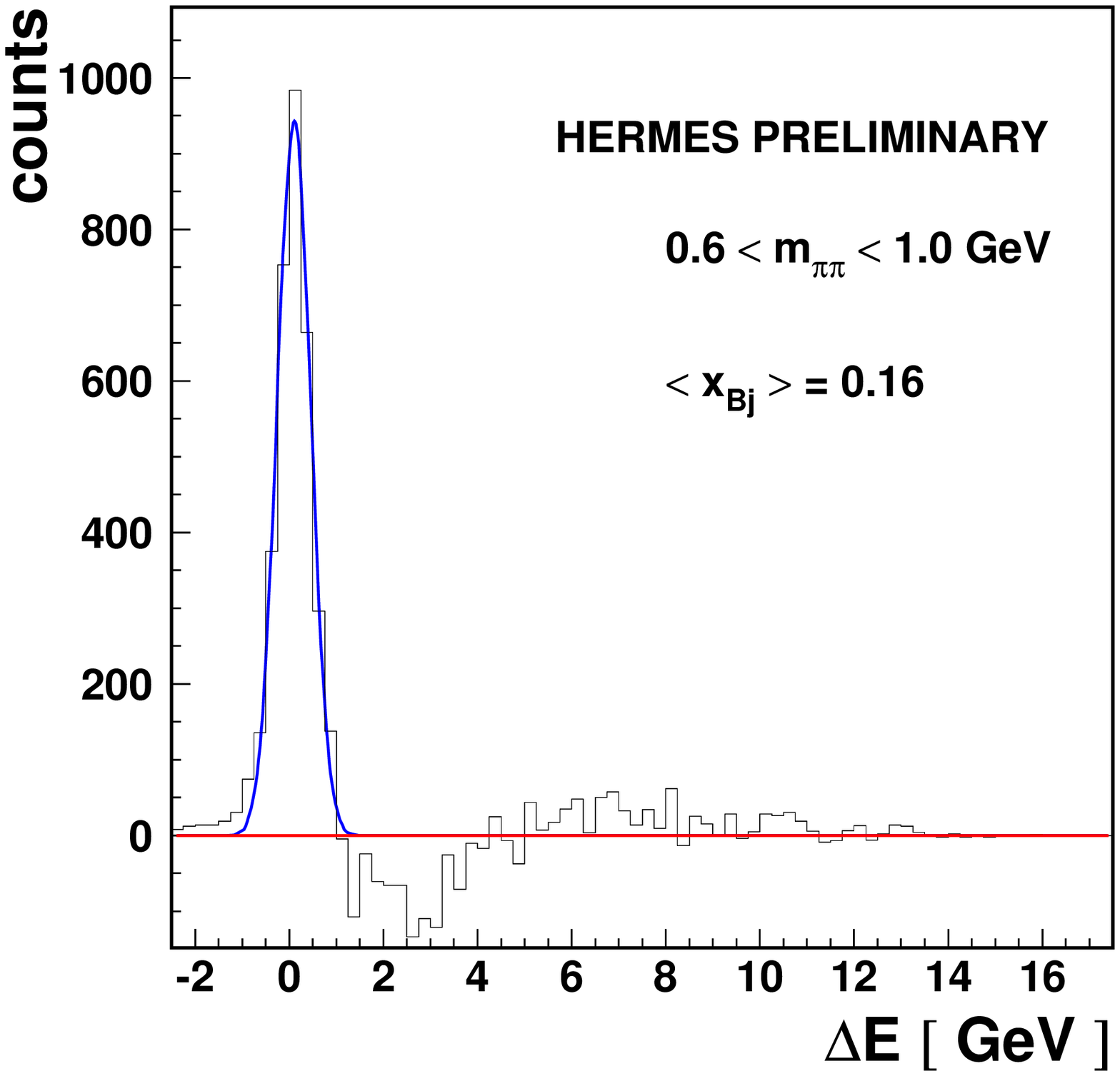}
  \caption{Left panel: {\it DIS MC} (shaded) normalized  
           to HERMES data (crosses) at $\Delta E > 2$ GeV. 
           Right panel: the exclusive $\pi^+\pi^-$ channel
           obtained by subtracting the {\it DIS}-MC 
           spectrum from the experimental one.
           In both panels only the $\rho^0$ 
           invariant mass window is considered.}
  \label{fig:norm}
\end{figure}

The most important source of background to the exclusive 
channel comes from fragmentation of partons in  
semi-inclusive deep inelastic scattering ({\it DIS}) 
at low $\Delta E$.
Due to the instrumental resolution and smearing, 
those events may contaminate the exclusive sample.
The shape of the {\it DIS} background is well   
reproduced by a $\it Monte \,Carlo$ simulation ({\it MC}). 
It is based on the {\it LEPTO} generator using the 
{\it LUND} model and the detector response was simulated 
by a {\it GEANT}-based $\it Monte \,Carlo$ code. 
%As can be seen in Fig.\ref{fig:norm} (left panel) 
{\it MC} data were normalized to experimental data at 
$\Delta E > 2$ GeV and then subtracted.
In order to illustrate our procedure, in
Fig. \ref{fig:norm} both the unsubtracted (left panel) 
and full subtracted (right panel) $\pi^+\pi^-$ data 
are shown as a function of the inelasticity $\Delta E$. 
Here only data within the $\pi^+\pi^-$ 
invariant mass window $0.6<m_{\pi\pi}<1.0$ GeV,   
around the $\rho^0$ mass, are shown.
Exclusive events were then selected for 
$\Delta E < 0.625$ GeV to maximize the ratio 
of the exclusive signal over the background  
($Sg/Bg$) and minimize its relative statistical error.
For these events the intensity density was calculated
in 10 bins of $m_{\pi \pi}$.
The intensity density of the background, 
$\langle cos\theta \rangle_{Bg}$, 
was evaluated using data for $\Delta E > 2$ 
GeV, assuming its $\Delta E$ independence.
This assumption was tested checking the
stability of $\langle cos\theta \rangle_{Bg}$
in different $\Delta E$ bins. 
In every bin of $m_{\pi \pi}$, the value of 
$\langle cos\theta \rangle_{Bg}$, weighted by 
the ratio $Sg/Bg$ for the chosen $\Delta E$ cut, 
was subtracted.
The background corrections range 
between $10-70\%$ in the various bins.
%
%--------------------------Results--------------------------%
%

\vspace{-0.7cm}
\section{Results}

\begin{figure}[b]
  \includegraphics[height=6cm, width=7.5cm]{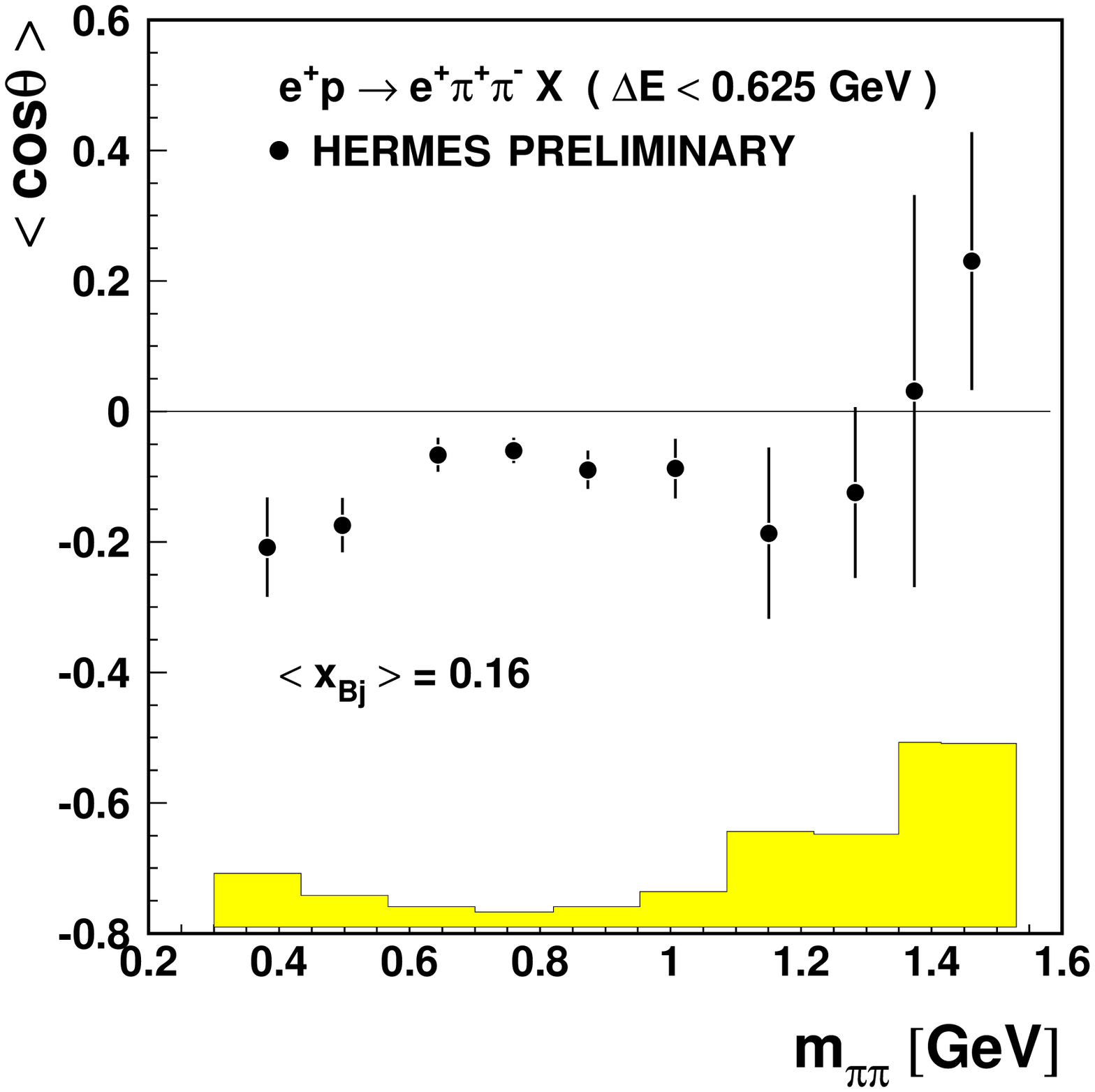}
  \includegraphics[height=6cm, width=7.5cm]{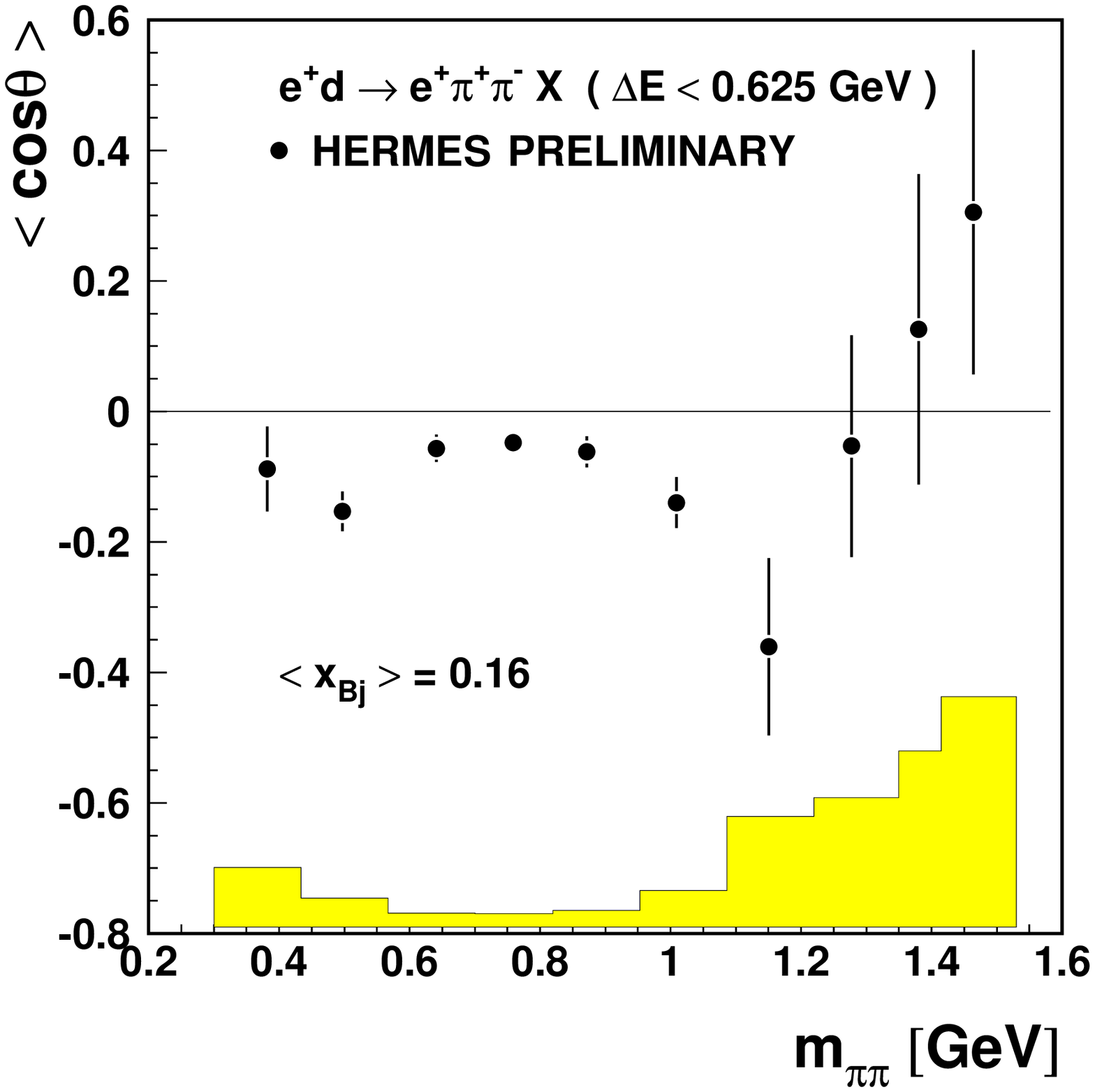}
  \caption{$m_{\pi\pi}$-dependence of the intensity 
           density \cosine for the proton (left panel)  
           and the deuteron (right panel).
           Shaded areas in both panels represent 
           the systematic uncertainty.}
  \label{fig:results_mhh}
\end{figure}

\vspace{-0.3cm}
In Fig. \ref{fig:results_mhh} the preliminary HERMES results 
on the $m_{\pi\pi}$-dependence of the intensity density 
\cosine are shown, for the proton in the left panel and for 
the deuteron in the right one.
The distributions show a clear angular asymmetry  
whose size changes with $m_{\pi\pi}$. 
At low $m_{\pi\pi}$ ($m_{\pi\pi}<0.6$ GeV) 
the asymmetry may be due to an   
interference between the lower tail of the $\rho^0$ meson 
and the non-resonant $\pi^+\pi^-$ $S$-wave production 
($I=1$ and $I=0$ interference). 
This interference is present over the entire 
invariant mass region considered. 
At large $m_{\pi\pi}$ ($m_{\pi\pi}>1.0$ GeV), additionally, 
an interference between the upper tail of the 
$\rho^0$ meson and the $f$-type mesons arises and is 
superimposed.
In particular, the possible change of sign of 
the asymmetry at $m_{\pi\pi}\approx 1.3$ GeV 
may be understood as being caused by the interference 
of the broad $\rho^0$ tail and the $f_2$ resonance 
($1.270$ GeV).
Note that no similar behavior is seen at the narrow
$f_0$ resonance ($0.980$ GeV), possibly due to the experimental 
resolution.

\begin{figure}[t]
  \includegraphics[height=5.5cm, width=8.cm]{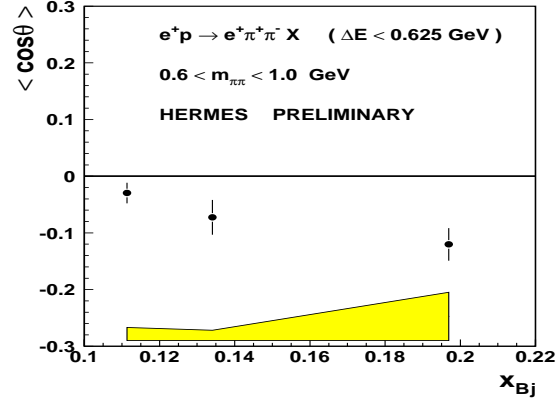}
  \caption{$x_{Bj}$-dependence of the intensity 
           density \cosine in the $\rho^0$ invariant mass 
           window for the proton.
           The shaded area represents
           the systematic uncertainty.}

  \vspace{-2.0cm}
  \label{fig:results_xbj}
\end{figure}

In Fig. \ref{fig:results_xbj} the $x_{Bj}$-dependence 
of the intensity density \cosine is shown for the proton 
within the above defined $\rho^0$ window.
The size of the \cosine asymmetry increases with 
$x_{Bj}$. 
This experimental finding is in agreement with 
theoretical expectations according to which  
at increasing $x_{Bj}$ the $\pi^+\pi^-$ production 
becomes increasingly dominated by $q\bar{q}$ exchange, 
leading to a sizable admixture 
of isoscalar and isovector pion pairs. 
As explained above this leads to an enhancement of 
the interference term.

The systematic uncertainties have been evaluated 
considering all hadrons as pions, using different 
exclusive cuts and applying  various procedures 
for the normalization of the {\it MC} generated background 
to the data. 
The main contribution has been found to originate 
from using different exclusive cuts.
\begin{figure}[b]
  \includegraphics[height=5.0cm, width=7.5cm]{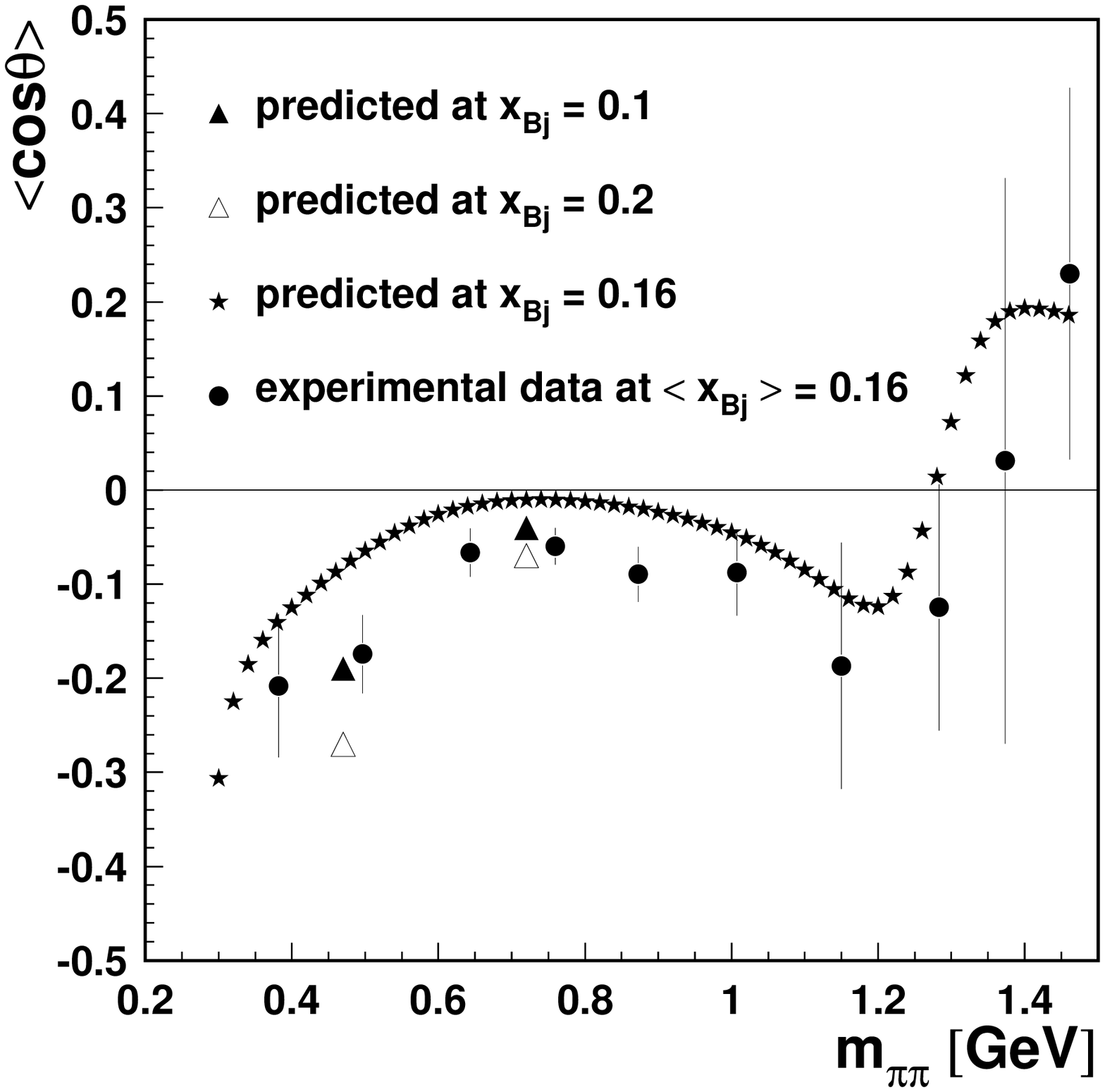}
  \includegraphics[height=5cm, width=7.5cm]{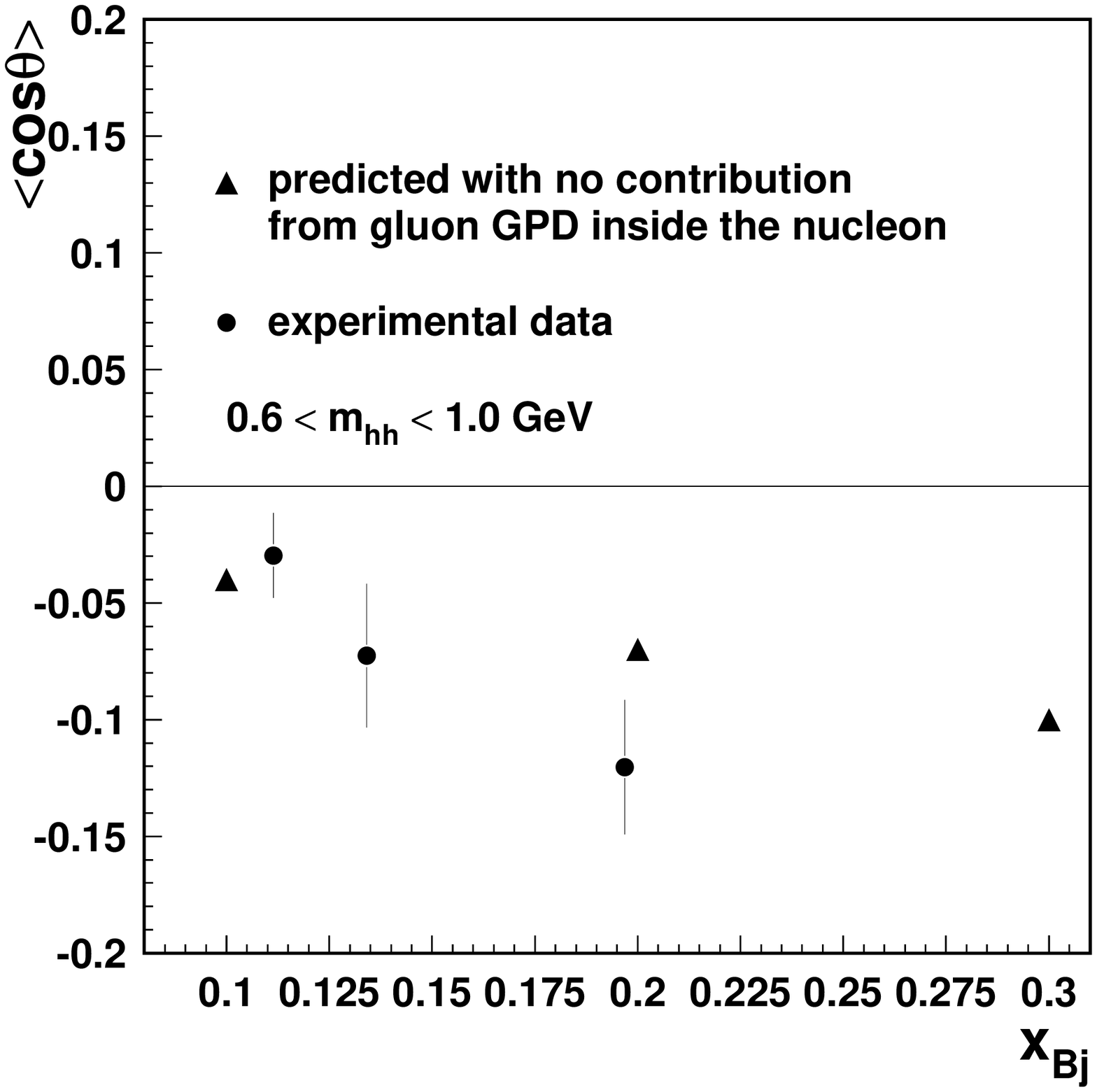}
  \caption{The experimental $m_{\pi\pi}$-dependence (left panel)
           and $x_{Bj}$-dependence (right panel) of the 
           intensity density \cosine for the proton 
           are compared with theoretical predictions in
           \cite{polyakov00,polyakov01}.
           {\it Triangles} ({\it stars}) show the predictions 
           with the gluon {\it GPD} neglected 
           \cite{polyakov00} (included \cite{polyakov01}).
           In the left panel, {\it Triangles} have been 
           slightly shifted for a better visibility.}
  \label{fig:theory}
\end{figure}
Using the diffractive {\it DIPSI MC} generator \cite{dipsi}, 
effects due to the acceptance were found to be negligible.
Radiative corrections were shown in \cite{akushevich} to be 
smaller than 1\% in the H1 and ZEUS kinematics.   
At larger $x_{Bj}$, where the HERMES analysis is performed, 
they are even smaller, and for that reason have been 
neglected.

The comparison with the theoretical predictions available 
so far is promising. 
In the left (right) panel of Fig. \ref{fig:theory} 
the experimental dependence of the intensity density 
\cosine on $m_{\pi\pi}$($x_{Bj}$) for the proton is compared 
with predictions performed at leading twist 
\cite{polyakov00,polyakov01}. 
The shape of the theoretical distribution is nicely 
reproducing the data, in particular at the $f_2$ meson 
mass where the asymmetry changes sign.
The reasonable agreement of the leading twist predictions 
with data may be understood as to arise from the cancellation 
of higher twist effects in this kind of asymmetry \cite{garcon}.

%%%%%%%%%%%%%%%%%%%%%%%%%%%%%%
%%        CONCLUSIONS       %% 
%%%%%%%%%%%%%%%%%%%%%%%%%%%%%%
\vspace{-0.6cm}
\section{Conclusions}

\vspace{-0.3cm}
The $l=1$ intensity density in $\pi^+\pi^-$ 
hard exclusive electroproduction was measured 
for the first time, for both proton and deuteron 
at $x_{Bj} > 0.1$. 
The quantity $\langle cos\theta \rangle$ 
is sensitive to the interference between the 
isoscalar channel ($I=0$) and the isovector channel 
($I=1$). 
The absolute value of the asymmetry measured in 
the $l=1$ intensity density \cosine shows a minimum at 
$m_{\pi\pi}=m_{\rho^0}$. 
At smaller invariant mass the asymmetry may be 
interpreted as originating from the interference 
of the $\rho^0$ lower tail with the non-resonant 
$\pi^+\pi^-$ production  ($S$-wave), and at larger 
invariant mass from the interference of the 
$\rho^0$ upper tail ($P$-wave) with $\pi^+\pi^-$ 
production from the $f_2$ ($D$-wave).
The interference signal in the $\rho^0$ 
invariant mass window was shown to increase in size 
with $x_{Bj}$. This behavior may be understood 
as to arise from the increased dominance of $q\bar{q}$ 
exchange at larger values of $x_{Bj}$, which 
leads to a sizable admixture of isoscalar 
and isovector pion pairs.
  
This is the first evidence of an $I=0$ admixture 
in $\pi^+ \pi^-$ exclusive electroproduction. 
This fact by itself is interesting in view of 
the importance of the scalar meson sector. 

In the {\it GPD} framework, theoretical predictions 
are available with and without 
inclusion of the {\it GPD} describing the two-gluon 
exchange. 
Both predictions appear to be consistent with the data. 
Therefore these results on exclusive $\pi^+ \pi^-$  
production, together with results from 
different exclusive channels (e.g. Deeply Virtual Compton 
Scattering \cite{airapetian01},  exclusive $\pi^+$ 
production \cite{airapetian02}) may lead   
to a better modeling of {\it GPD}s.

%%%%%%%%%%%%%%%%%%%%%%%%%%%%%%%%%%%%%%%%%%%%%%%%
%% ACKNOWLEDGMENTS
%%%%%%%%%%%%%%%%%%%%%%%%%%%%%%%%%%%%%%%%%%%%%%%%

\vspace{-0.5cm}
\begin{theacknowledgments}

\vspace{-0.3cm}
  \small
  We are deeply grateful to N. Bianchi, A. Borissov, 
  M. Diehl, W.-D. Nowak, B. Pire and M.V. Polyakov for useful 
  discussions and suggestions.
\end{theacknowledgments}
%%%%%%%%%%%%%%%%%%%%% BIBLIOGRAFY %%%%%%%%%%%%%%%%%%%%%%%

\end{document}